\documentclass[traditabstract]{aa} 
\usepackage{amsmath} 
\usepackage[hyphens]{url} 
\usepackage{color}
\usepackage{natbib}
\bibpunct{(}{)}{;}{a}{}{,}    

\usepackage{graphicx}
\usepackage{txfonts}

\begin{document}
   \title{A CRIRES-search for H$_{3}^{+}$ emission {from the hot Jupiter atmosphere of HD~209458~b}\thanks{Based on observations collected at the European Organisation for Astronomical Research in the Southern Hemisphere, Chile, 086.C-0045.}}
\titlerunning{H$_{3}^{+}$ emission limits with CRIRES}


   \author{L. F. Lenz,
          \inst{1}
          A. Reiners,
	  \inst{1}
	  A. Seifahrt,
	  \inst{2}
	  \and
	  H. U. K\"aufl
	  \inst{3}}

   \institute{Institute for Astrophysics, University of Goettingen,
              Friedrich-Hund-Platz 1, 37077 Goettingen\\
              \email{lealenz@astro.physik.uni-goettingen.de}
         \and
            Department of Astronomy and Astrophysics,
University of Chicago, IL 60637, USA
           \and
           European Southern Observatory (ESO), Karl-Schwarzschild-Str. 2, 85748 Garching, Germany\\
             }

   \date{Received [...]; accepted [...]}

 
  \abstract
  {Close-in extrasolar giant planets are expected to cool their thermospheres by producing H$_{3}^{+}$ emission in the near-infrared (NIR), but simulations predict  H$_{3}^{+}$ emission intensities that differ in the resulting intensity by several orders of magnitude. 
    We want to test the observability of H$_{3}^{+}$ emission with CRIRES at the Very Large Telescope (VLT), providing adequate spectral resolution for planetary atmospheric lines in NIR spectra. We search for signatures of planetary H$_{3}^{+}$  emission in the $\mathrm{L}'$ band, using spectra of HD~209458 obtained during and after secondary eclipse of its transiting planet HD~209458~b. We searched for  H$_{3}^{+}$ emission signatures in spectra containing the combined light of the star and, possibly, the planet. With the information on the ephemeris of the transiting planet, we derive the radial velocities at the time of observation and search for the emission at the expected line positions. We also apply a cross-correlation test to search for planetary signals and use a shift \& add technique combining all observed spectra taken after secondary eclipse to calculate an upper emission limit.
   We do not find signatures of atmospheric H$_{3}^{+}$ emission in the spectra containing the combined light of HD~209458 and its orbiting planet. We calculate the emission limit for the H$_{3}^{+}$ line at 3953.0~nm $[Q(1,0)]$ to be $8.32\cdot 10^{18}\,\mathrm{W}$ and a limit of $5.34\cdot 10^{18}\,\mathrm{W}$ for the line at 3985.5~nm $[Q(3,0)]$.   
   Comparing our emission limits to the theoretical predictions suggests that we lack 1 to 3 magnitudes of sensitivity to measure H$_{3}^{+}$ emission in our target object. We show that under more favorable weather conditions the data quality can be improved significantly, reaching $5 \cdot 10^{16}\,\mathrm{W}$ for star-planet systems that are close to Earth. We estimate that pushing the detection limit down to $ 10^{15}\,\mathrm{W}$ will be possible with ground-based observations with future instrumentation, for example, the European Extremly Large Telescope. } 

\maketitle
\section{Introduction}
Methods of exoplanet detection have become more and more successful in recent years and various kinds of exoplanets, from hot Jupiters to earth-sized planets, have recently been detected. However, it is still difficult to gain information about the atmospheric composition of these exoplanets. Only a small fraction of the detected exoplanets can be accessed for investigations of the atmosphere, for example, transits and secondary eclipses.

HD\,209458\,b was the first transiting exoplanet for which an atmosphere was detected in a spectrophotometric observation of  \ion{Na}{I} \cite[]{Charbonneau2002}.  Most exoplanets with a detection of atmospheric molecules are gas giant planets in close-in orbits.
The atmospheres of  these extrasolar giant planets (EGP), such as HD\,209458\,b, are expected to be severely influenced by the stellar extreme ultraviolet (EUV) radiation that drives a hydrodynamic escape in the upper atmosphere. Different models have been derived to estimate the evaporation rate of close-in giant planets \cite[e.g.,][]{Lammer2003}.  

Some of these models propose that infrared (IR) emissions of H$_{3}^{+}$ ions cool the thermospheres of close-in EGPs and, thus, contribute to balance the heating due to the host star radiation \cite[]{Yelle2004}. H$_{3}^{+}$ emission was detected from the auroral regions of Jupiter by \cite{Drossart1989}. The ion was successfully used as a tool to measure atmospheric temperatures and ion densities of Jupiters atmosphere  \cite[]{Yelle2004}. \cite{Miller2000} argue this ion is the main coolant in the  ionosphere and thermosphere of Jupiter in both the auroral and nonauroral regions. H$_{3}^{+}$ was also detected in Saturn \cite[]{Geballe1993} and Uranus \cite[]{Trafton1993}.  The temperature for the thermosphere of Saturn is measured at around $800\,\mathrm{K}$  \cite[]{Geballe1993}, and for Jupiter a temperature range of approximately $700-1,000\,\mathrm{K}$ was measured by \cite{Lam1997}.  \cite{Koskinen2010a} measure a mean temperature of $8,000-11,000\,\mathrm{K}$ for the thermosphere of HD209458b.  H$_{3}^{+}$ is an effective coolant even for high temperatures up to $10,000\,\mathrm{K}$\cite[]{Neale1996}.

 H$_{3}^{+}$ is formed by a reaction of H$_{2}^{+}$, under extreme ultraviolet (EUV) radiation and energetic electron precipitation along magnetic field lines, reacting with neutral H$_{2}$, i.e.,

  \begin{align*}
 \mathrm{H}_{2} + \mathrm{e}^* & \rightarrow   \mathrm{H}_{2}^{+}  + \mathrm{e} + \mathrm{e} \\  
    \mathrm{H}_{2} + h \nu & \rightarrow   \mathrm{H}_{2}^{+}  + \mathrm{e}  \\
  \mathrm{H}_{2} + \mathrm{H}_{2}^{+}  &\rightarrow  \mathrm{H}_{3}^{+} + \mathrm{H} .
 \end{align*}

  The ion is destroyed by the dissociative recombinations
   \begin{align*}
  \mathrm{H}_{3}^{+} + \mathrm{e}   &\rightarrow \mathrm{H}_{2}^{+} +  \mathrm{H}  + \mathrm{e} \\
   \mathrm{H}_{3}^{+} + \mathrm{e} &\rightarrow \mathrm{H}  + \mathrm{H}  + \mathrm{H} .
 \end{align*}

  So far  H$_{3}^{+}$  emission has not been detected in the atmosphere of an EGP on a planet outside of our solar system. A detection would help to understand the thermal structure of EGPs. Additionally, the planetary radial velocity could be measured directly: since  H$_{3}^{+}$ is not formed in stellar atmospheres, a detection could safely be assigned to the exoplanet of the observed stellar system. 
  
The theoretical predictions for H$_{3}^{+}$ emitted from exoplanet atmospheres vary by several magnitudes in the different models. 
\citet{Miller2000} estimated an emission limit of  $\sim 10^{17}\,\mathrm{W}$ for a planet similar to $\tau$Boo~b. 
  \citet{Yelle2004} investigated the case of HD\,209458\,b with a one-dimensional model of an EGP atmosphere. He calculated an emission limit of $1\cdot 10^{16}\,\mathrm{W}$ emitted from the lower thermosphere. At higher altitudes he derived temperatures of $10,000$ to $15,000\,\mathrm{K}$, 
  where H$^{+}$ becomes dominant and suppresses the formation of H$_{3}^{+}$ . 
  
 The lowest estimations for the H$_{3}^{+}$ emission limits of close-in EGPs are derived by \cite{Koskinen2007}: With their model of a coupled thermosphere and ionosphere they performed three-dimensional, self-consistent global simulations for different orbital distances of EGPs around a sunlike host star. They conclude, that EGPs are cooled efficiently by H$_{3}^{+}$ inside $0.2-1\,\mathrm{AU}$ orbits and state that thermal dissociation and dissociative photoionization of H$_{2}$ hampers the emission for orbits closer than $0.1\,\mathrm{AU}$. With their simulation they derived a total power output of 
$\sim 10^{15}\,\mathrm{W}$ for a hot Jupiter planet in a $0.1\,\mathrm{AU}$ orbit and calculated a spectral line output  of $\sim 10^{12}\,\mathrm{W}$ for the intensity of the $Q(3,0)$-transition.

\cite{Brittain2002} reported the detection of  H$_{3}^{+}$ emission signals from observations of HD~141569~A. However these observations could not be confirmed by observations made by \cite{Goto2005}, who measured upper emission limits that were significantly lower than the signals reported by \cite{Brittain2002} while achieving comparable data quality.
The most extensive observation attempt for H$_{3}^{+}$ emission from hot Jupiter systems was carried out by \cite{Shkolnik2006}. They observed six close-in EGPs with CSHELL at the NASA IRTF. They searched for emission from the $Q(1,0)$-transition at 3953.0\,nm and calculated emission limits around $1\cdot10^{18}\,\mathrm{W}$  from their measurements. The lowest limit is reached for GJ~436 with $6.3\cdot10^{17}\,\mathrm{W}$.  \cite{Laughlin2008} reported an emission limit of $9.0\cdot10^{17}\,\mathrm{W}$ for observations of $\tau$Boo with CSHELL.
\cite{Maillard2011} suggested an observation strategy using high spectral resolution and possibly occultation spectroscopy in order to differentiate planetary and stellar flux. They stated that the growing sample of known exoplanets also offers more possible candidates for the search for H$_{3}^{+}$ emission.

This work investigates the feasibility of H$_{3}^{+}$ observations with CRIRES \cite[]{2004SPIE.5492.1218K} at the VLT. The CRIRES spectrograph provides the highest resolution in the infrared that is available today. 

\section{Observations}
\label{sec:Obs}
The observations were carried out with CRIRES \cite[]{2004SPIE.5492.1218K} at the VLT. The spectra were taken with the reference wavelength set to 4010~nm in the $\mathrm{L}'$~band with a resolving power of R$\sim$100\,000.
\begin{table}
\caption{Wavelength coverage of the observed spectra for each of the four CRIRES detectors and the H$_{3}^{+}$ emission line positions and intensities from the southern Aurorae of Jupiter as given in \citet{Maillard1990}.  }             
\label{tab:setting}      
\centering                          
\begin{tabular}{l c c c }        
\hline\hline                 
Detector & Wavelength range  &           Line position   &              Intensity                         \\ 
                  &             $\mathrm{[nm]}$ & $\mathrm{[nm]}$        &        $[\mathrm{\frac{W}{cm^{2} sr}}]$ \\ \hline   
1 &     3947.3 - 3968.6 & 3953.0        &        42.8    \\
2 &     3974.6 - 3995.1 & 3985.5        &       45.2    \\
   &                             & 3987.0   &   23.5                                    \\
   &                             & 3994.6       &       11.5                    \\
3 &     4000.6 - 4020.3 & 4012.0        &       19.3    \\
 &                               & 4013.3       &       17.3                                            \\
4 &     4025.4 - 4044.3 & 4043.2        &       10.2    \\
\hline                                   
\end{tabular}
\end{table}

Our chosen setting covers seven H$_{3}^{+}$ lines that were found in strong emission in the  southern auroral zone of Jupiter by \citet{Maillard1990}. Three of the emission lines of the atmosphere of Jupiter were measured with an intensity larger than $20\,\mathrm {W cm^{-2} sr^{-1}}$ and another four emission lines with an intensity above $10\,\mathrm {W cm^{-2} sr^{-1}}$. 

The line positions are listed in Table~\ref{tab:setting} with the corresponding intensities from \citet{Maillard1990} and the detector number and wavelength coverage of the four CRIRES detectors, on which the lines fall  in the observed spectra.
The H$_{3}^{+}$ emission line at 3953.0\,nm was also used by \citet{Shkolnik2006} in their search for H$_{3}^{+}$ of hot Jupiter atmospheres.

\begin{table}
\caption{System parameters \cite[]{Southworth2010}.}             
\label{tab:systemparameters}      
\centering                          
\begin{tabular}{l c c c c c }        
\hline\hline                 

Object & Spectral Type  & Distance       & M$_{\mathrm{Star}}$   & M$_{\mathrm{P}}$                 & P$_{\mathrm{orb}}$  \\
                &                       &  [pc]         &  [M$_{\odot}$]                 & [M$_{\mathrm{Jup}}$]  &  [d]  \\ \hline
HD\,209458 & G0                 &       47.1            & 1.01                          & 0.685 &                         3.5247 \\
\hline                                   
\end{tabular}
\end{table}

As a result of unfavorable weather conditions, we were able to record  only 16 spectra of HD~209458. The stellar and planetary parameters of HD~209458 are summarized in Table~\ref{tab:systemparameters}. Twelve of the observed spectra were taken during secondary eclipse. Hence, planetary emission might be visible in the four remaining spectra available for  analysis; these four spectra are hereafter called combined light spectra. The mean signal-to-noise ratio with exposure times of $150\,\mathrm{s}$ is $\approx71$ for the 12 eclipse spectra and for the four combined light spectra.

\begin{figure*}
   \centering
   \includegraphics[width=0.42\textwidth]{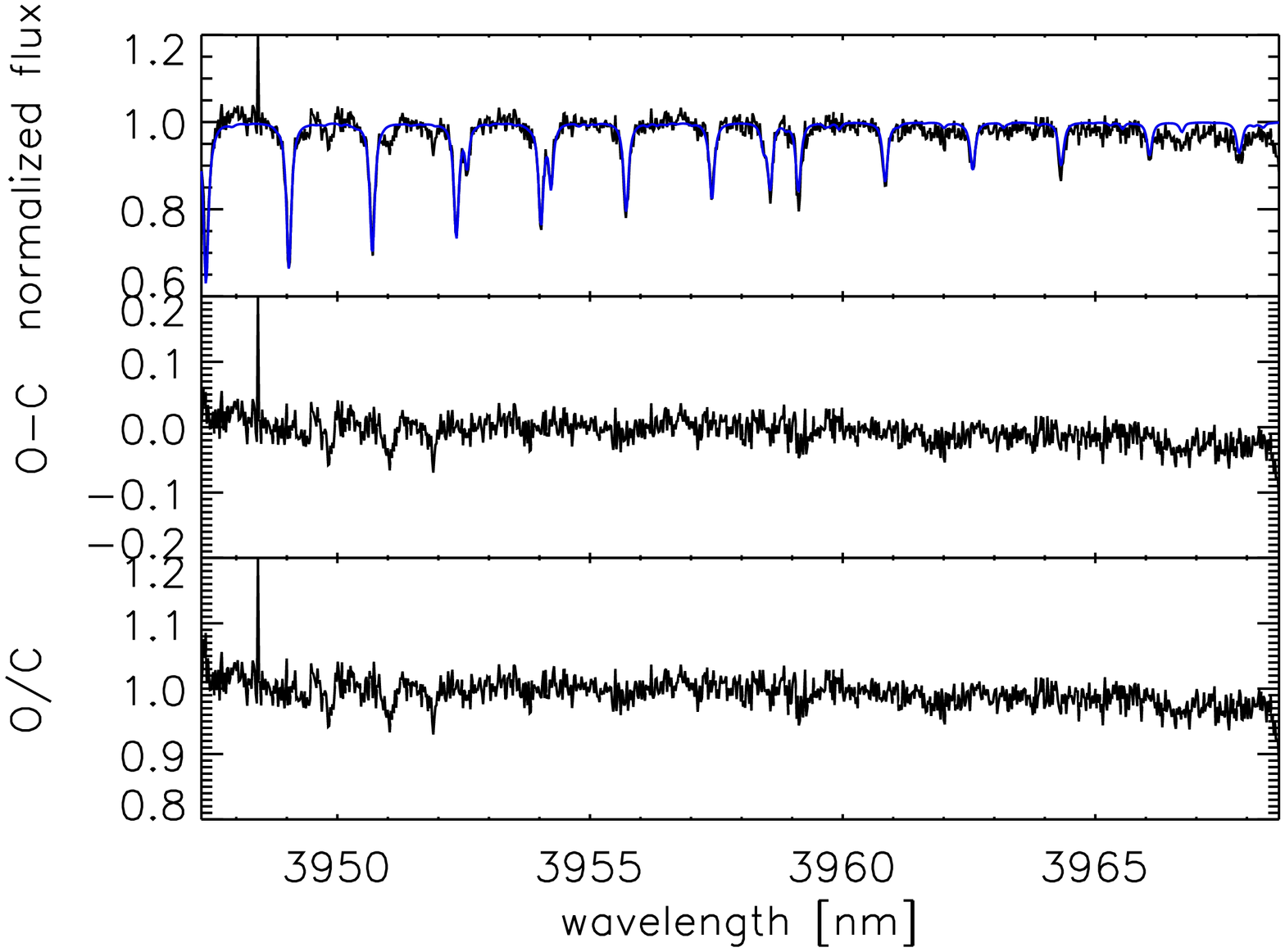}
   \includegraphics[width=0.42\textwidth]{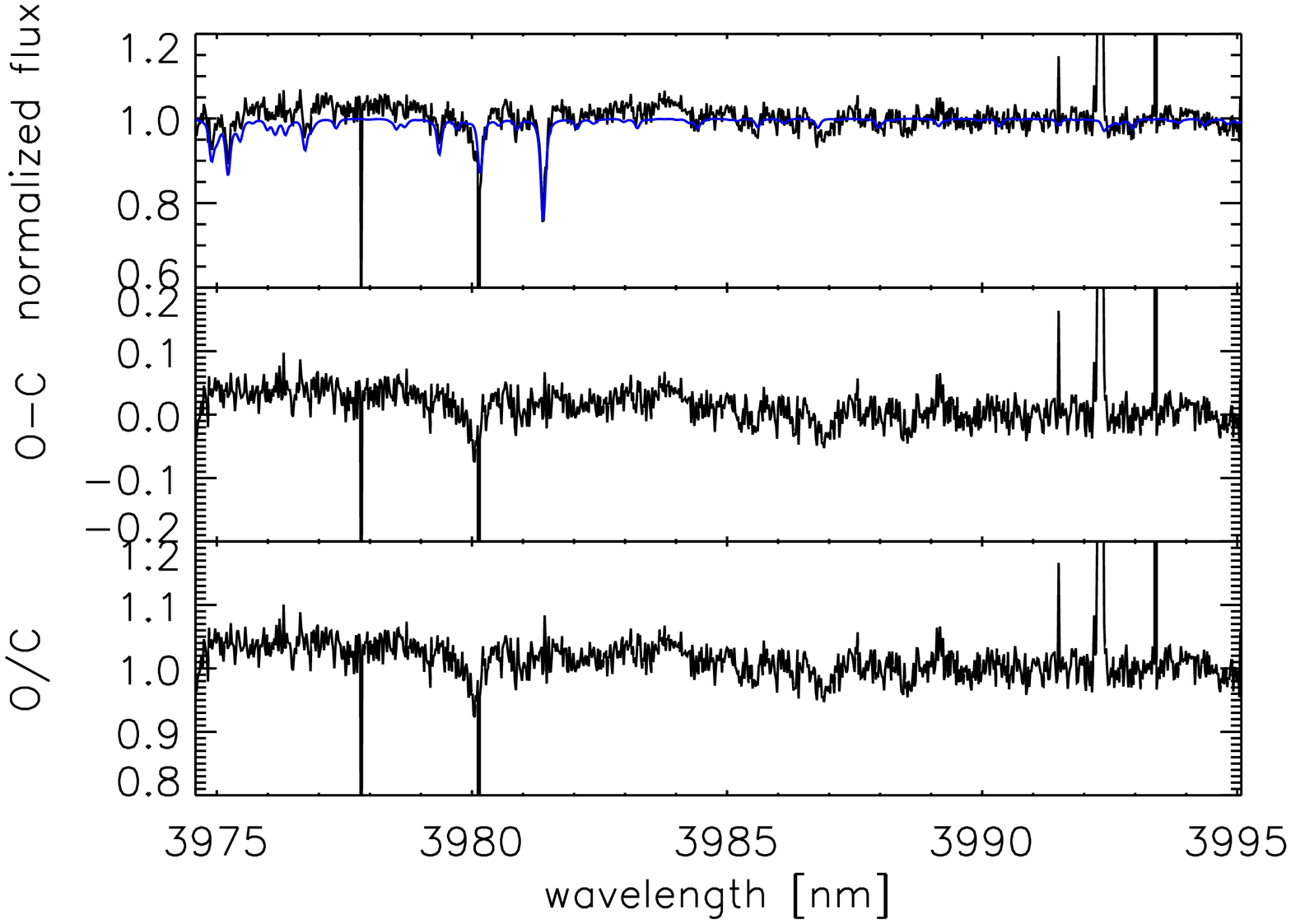}
   
      \includegraphics[width=0.42\textwidth]{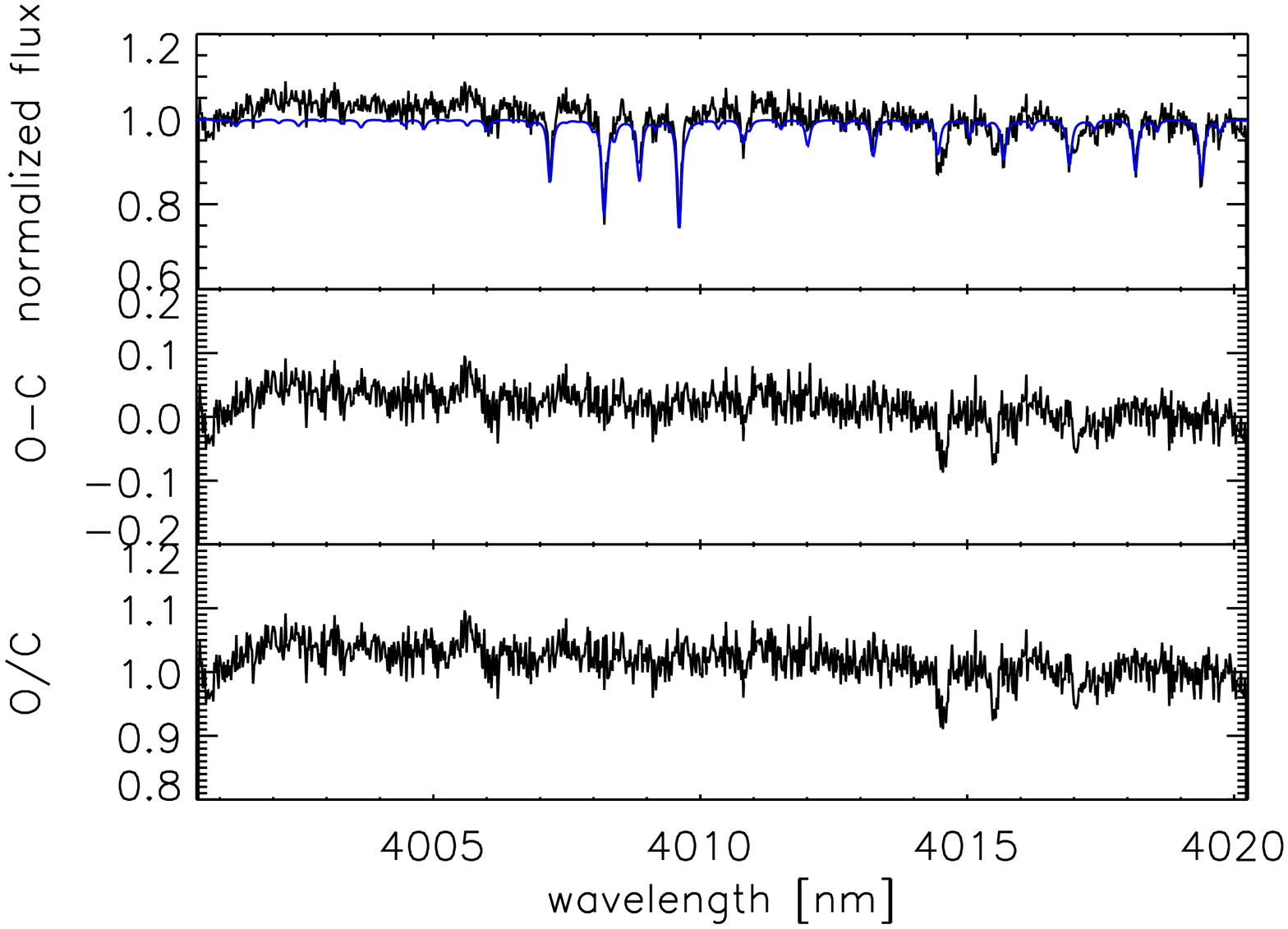}
   \includegraphics[width=0.42\textwidth]{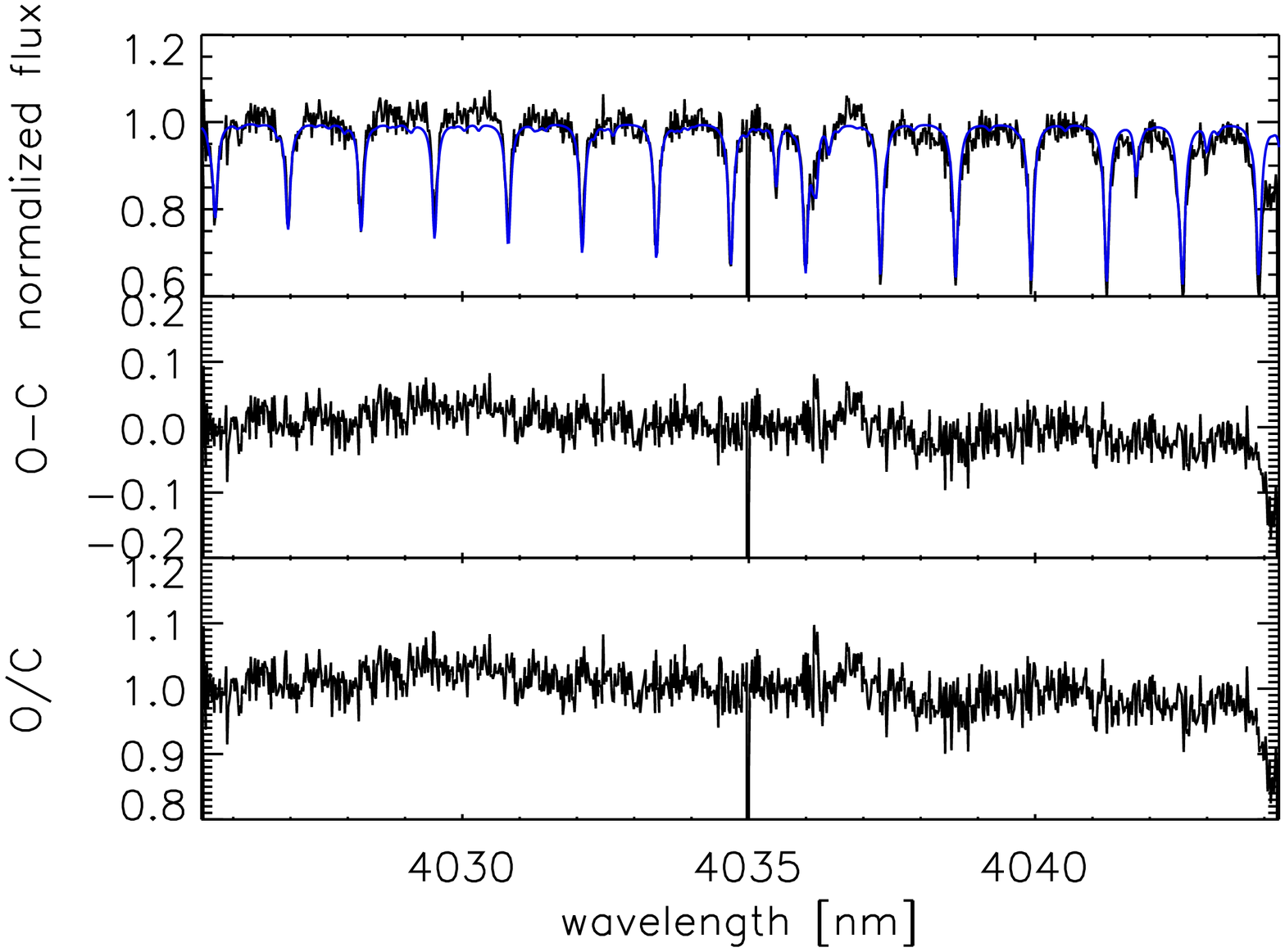}
    \caption{Example spectrum of HD\,209458 for the 4010\,nm setting of CRIRES, showing all four detectors. \textit{Upper panel:} CRIRES spectrum (black) and synthetic telluric spectrum (blue) from the telluric modeling. \textit{Lower panels:} observed-computed (O-C) residuals and observed/computed (O/C) results. The O/C spectra are used in the analysis. }
       \label{fig:bothspec}

   \end{figure*}

The observations for HD\,209458 were carried out in an ABBA-nodding pattern.
The data reduction was performed using the CRIRES pipeline recipes (Version 2.1.1) provided by ESO\footnote{\url{http://www.eso.org/sci/software/pipelines/}}. The optimal 
extraction of the spectra was performed separately for the A  and B frames to avoid a degradation of the spectral resolution due to 
the curvature of the slit.
To correct for atmospheric absorption features and solve for the wavelength solution of the spectra, atmospheric modeling was applied, using the 
LBLRTM\footnote{\url{http://rtweb.aer.com/lblrtm_description.html}} code with the techniques described in \cite{Seifahrt2010}. A typical spectrum (black) with its corresponding telluric spectrum from the modeling (blue) is shown in the upper panel of Fig.~\ref{fig:bothspec}.  In the lower panels the observed-computed (O-C) residuals and observed/computed (O/C) results are shown.  To correct for  telluric absorption features, the observed spectra are divided by the model spectrum (O/C). 

In-dispersion stray light from the CRIRES grating causes $1-3\%$ residual flux in the absorption cores of the observed line profiles for regions with fully saturated telluric features \cite[]{Lebzelter2012}. For pointlike sources such as stars, this effect cannot yet be corrected for. 
Besides this effect, other systematic deviations caused by the telluric correction with the synthetic model are not expected.

The four detectors have a different continuum level (see Fig.~\ref{fig:bothspec}). We assume this trend to originate from the reduction of the flatfields. The spectra were treated with the standard flatfielding recipes for CRIRES provided by ESO.

We use the information on the time of observation, location of the observatory, and the stellar coordinates to calculate the topocentric velocities for the observed spectra and apply the correction to all spectra.

\begin{figure}
 \centering
   \includegraphics[width=0.5\textwidth]{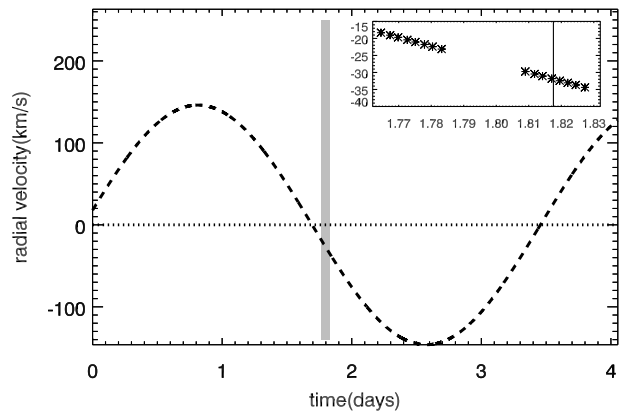}
       \caption{Radial velocity of HD\,209458\,b (dashed line). The shaded area indicates the timespan of the observations. The inset shows the different observations with the vertical line indicating the end of the secondary eclipse with the ephemeris value from \cite{Knutson2007} and the corresponding transit duration from \cite{Beaulieu2010}. Only observations after this point contain the combined light of the planet and host star. }
       \label{RV}
   \end{figure}

HD\,209458\,b is a transiting planet and we use the ephemeris value derived by \cite{Knutson2007} to calculate the planetary 
radial velocity value at the time of each observation. We derive the end of the secondary eclipse from the transit duration given by \cite{Beaulieu2010}.
Table~\ref{tMJRV} gives the observation times in MJD format and the calculated radial velocities of the planet for the combined light spectra. Figure\,\ref{RV} shows the radial velocity curve of the orbiting plane. The shaded area indicates the timespan of our observations.  The inset shows each separate observation, with the vertical line indicating the end of the secondary eclipse.
The uncertainties of the radial velocity resulting from the error of the ephemeris value given by  \cite{Knutson2007}  are negligible in comparison to the uncertainties of the ephemeris itself. Varying ephemeris values are reported for HD~209458\,b and the number of spectra in secondary transit depends on the ephemeris value used. Using, for example, the ephemeris value by \cite{Wittenmyer2005} instead of the  result by \cite{Knutson2007}, leads to five spectra where the planet is visible. 
We choose to use the most recent value derived by  \cite{Knutson2007},  and its  corresponding transit duration from \citet[]{Beaulieu2010}.

\begin{table}
\caption{MJD and planetary radial velocities of the combined light spectra.}             
\label{tMJRV}      
\centering                          
\begin{tabular}{l c c  }        
\hline\hline                 
\# spectra      & MJD & RV planet  \\    
        &               & $\mathrm{[kms^{-1}]}$         \\ \hline
1       &        55479.058888248        &       -32.42          \\ 
2       &       55479.061550962 & -33.04  \\
3       &       55479.064064636 &  -33.67 \\
4       &       55479.066697450         & -34.38 \\
\hline                                   
\end{tabular}
\end{table}

\section{Search for H$_{3}^{+}$ emission }
\label{sec:ana}
\subsection{Direct search}
We start searching for H$_{3}^{+}$ emission peaks in each of the combined light spectra separately, considering the radial velocity shift of the planet with respect to its host star. We expect that an emission peak would most probably occur in all combined light spectra, since they were observed within minutes of each other. Since HD\,209458 is tidally locked, emission lines from its atmospheres would be very narrow. In the direct search in all four combined light spectra, we cannot detect any signatures of  H$_{3}^{+}$ emission lines.

\subsection{Cross-correlation approach}
Our second approach to the search for emission features of the planetary atmosphere is to cross-correlate the combined light spectra pairwise. 
Emission lines of the planet would show the differential planetary radial velocity in the cross-correlation. Thus, a detection in a cross-correlation would be independent of the accuracy of the knowledge of the H$_{3}^{+}$ emission line positions. 
The detectors are treated separately because of the gaps between the detectors. Also, a $\sim0.06\,\mathrm{nm}$ chunk is cut from both ends of each spectrum to avoid signals from nonlinearity effects at the edges of the spectra. As a preparation for the cross-correlation test, all spectra from detector No. 2 are prepared in the following way: The crack and the right end of the detector is removed.  Strong spikes in all spectra are removed and replaced by a median smoothed spectrum. 

We apply the cross-correlation to the following combinations of spectra:
First, we use the telluric-corrected spectra for the cross-correlation. In the second step, we subtract the host star from the spectra, by computing a mean spectrum of the 12 secondary eclipse spectra and subtract the result from each combined light spectrum before performing the cross-correlations. 
In the third step, we subtract a single secondary eclipse spectrum instead of the mean eclipse spectrum. This single spectrum has a lower signal-to-noise ratio, but since the timespan between the last observed secondary eclipse spectrum and the combined light spectra is very short, parts of the atmospheric distortions that have not been corrected by the telluric modeling  vanish with the subtraction. However, if the ephemeris is not precise one might also subtract  a viable signal from the planet. We choose the secondary eclipse spectrum that was observed directly before the end of the secondary eclipse.
 As a consistency check, we repeat this procedure using the spectrum that was observed second to last during secondary eclipse.

A planetary signal in the spectra is expected to cause a signal in the cross-correlation with the differential planetary radial velocity of the two spectra that are used as input spectra in the cross-correlation. 
Since the spectra were recorded in close succession shortly after the secondary eclipse, the differential planetary radial velocity values are only up to $2\,\mathrm{km\,s^{-1}}$. Also, the stellar and telluric lines, as well as detector artifacts, cause a peak around $0\,\mathrm{km\,s^{-1}}$.

The cross-correlation results for the pure spectra and the spectra with subtracted mean eclipse spectra show various peaks over the whole radial velocity range. The results for the spectra with subtracted last eclipse spectrum show a clear signal around  $0\,\mathrm{km\,s^{-1}}$ as well as the results for the consistency check for which another eclipse spectrum is subtracted before the cross-correlation is applied. We continue the analysis with the spectra from which we subtracted the last spectrum that was observed during secondary eclipse.

\begin{figure}
   \centering
   \includegraphics[width=0.45\textwidth]{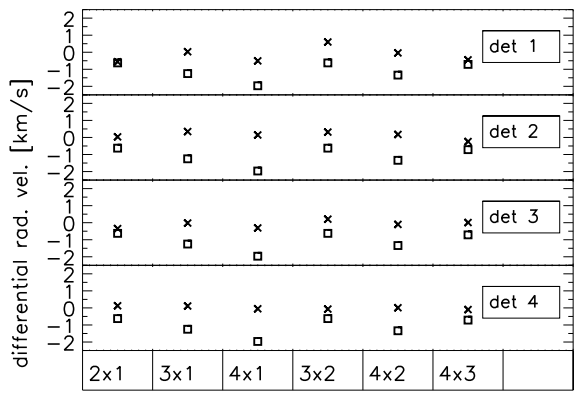}
       \caption{Cross-correlation results. Upper to lower panels show the four CRIRES detectors. The spectrum that was observed last during secondary eclipse is subtracted from each combined light spectrum. The resulting spectra are cross-correlated and the main peak is fitted with a Gaussian. We derive the radial velocity value from the maximum of the fit and expect a planetary signal to shift the Gaussian fit sideways, producing a signal between  $0\,\mathrm{km\,s^{-1}}$ and the differential radial velocity in case of a detection.  The squares indicate the calculated differential radial velocity values from the time of observation and ephemeris and the crosses show the measured radial velocity from the cross-correlation. The error bars lie within the symbol size. }
       \label{crossresults}
   \end{figure}

We fit the main peak of each cross-correlation with a Gaussian and derive the radial velocity from the maximum of the fit. We expect a planetary signal in the cross-correlation to shift the Gaussian fit sideways. The fitted velocity would thus lie between  $0\,\mathrm{km\,s^{-1}}$ and the differential radial velocity in case of a detection.
In Fig.~\ref{crossresults} the results of the cross-correlation are presented for all four detectors. The x-axis gives the number of spectra used for the cross-correlations (e.g., 1x2 means the cross-correlation of the 1st and 2nd observed combined light spectra). 
The squares show the differential planetary radial velocities of the pair of spectra and the crosses give the fitted radial velocity values from the cross-correlations. 
In case of a detection of planetary emission, we expect the radial velocity values of the fit (squares) to follow the pattern of the differential radial velocity values (crosses) in the plot. We expect a planetary signal to occur in all observed spectra, since they were recorded in close succession.

We cannot find cross-correlation signals that follow the expected pattern of the planetary radial velocity. Fitting the cross-correlation results give signals that scatter around $0\,\mathrm{km\,s^{-1}}$. 
The same result follows for the consistency test in which another secondary eclipse spectrum is subtracted from the combined light spectra. We also check all cross-correlation results by eye for peaks and do not detect any signal that indicates planetary emission.

\subsection{ Search for emission with shift and add}
The known ephemeris also enables us to use a shift and add approach to search for H$_{3}^{+}$ emission:
all observed combined light spectra are shifted in wavelength with respect to their apparent radial velocity of the planet at the time of observation. Afterward, the flux is summed up binwise. With this technique the stellar lines and residuals from the telluric correction get blurred and a planetary emission line that is hidden in the stellar flux would emerge in the summation of the shifted spectra, given that the ephemeris and consequent radial velocity values are correct.

\begin{figure*}
   \centering
   \includegraphics[width=0.42\textwidth]{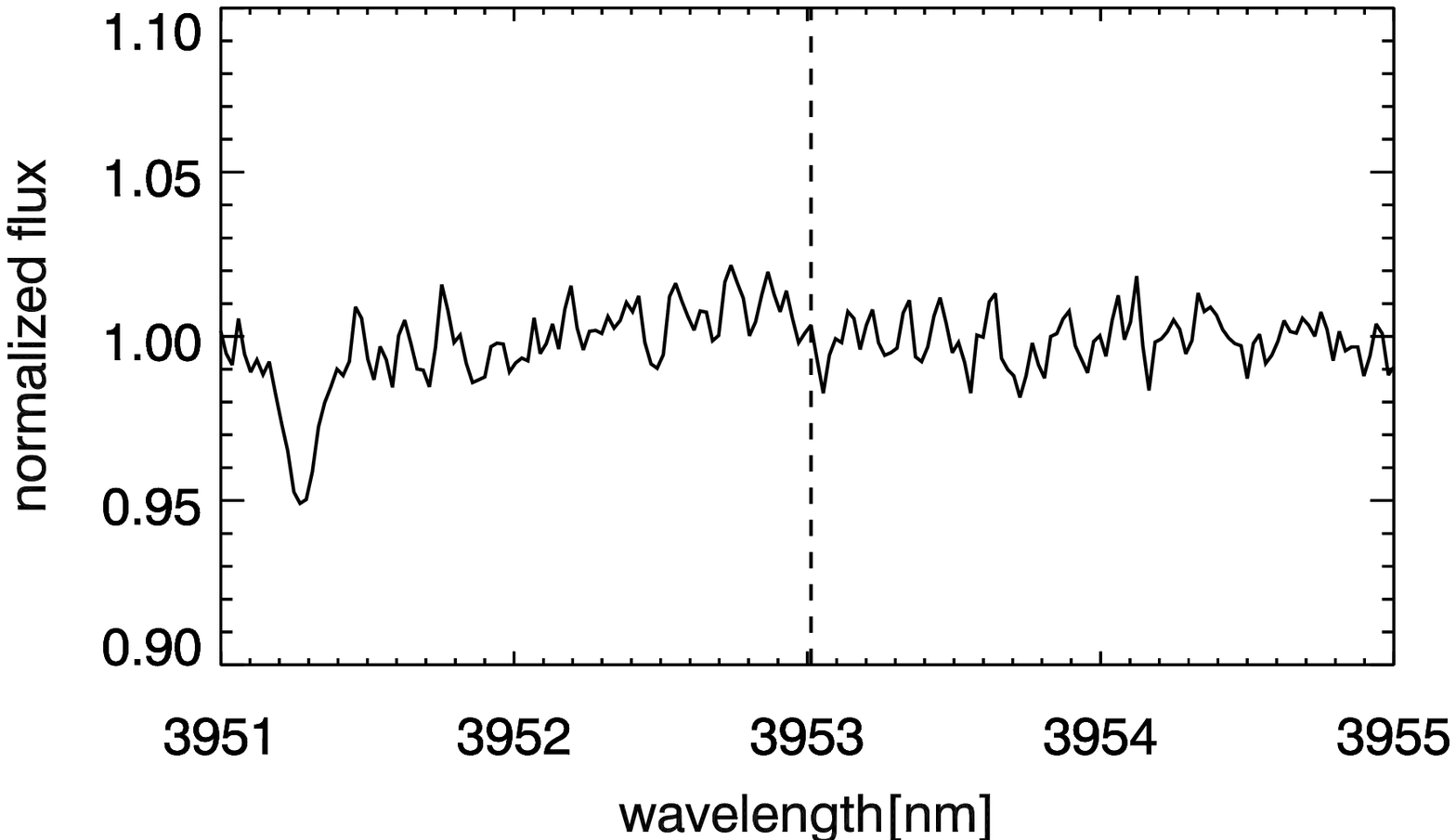}
   \includegraphics[width=0.42\textwidth]{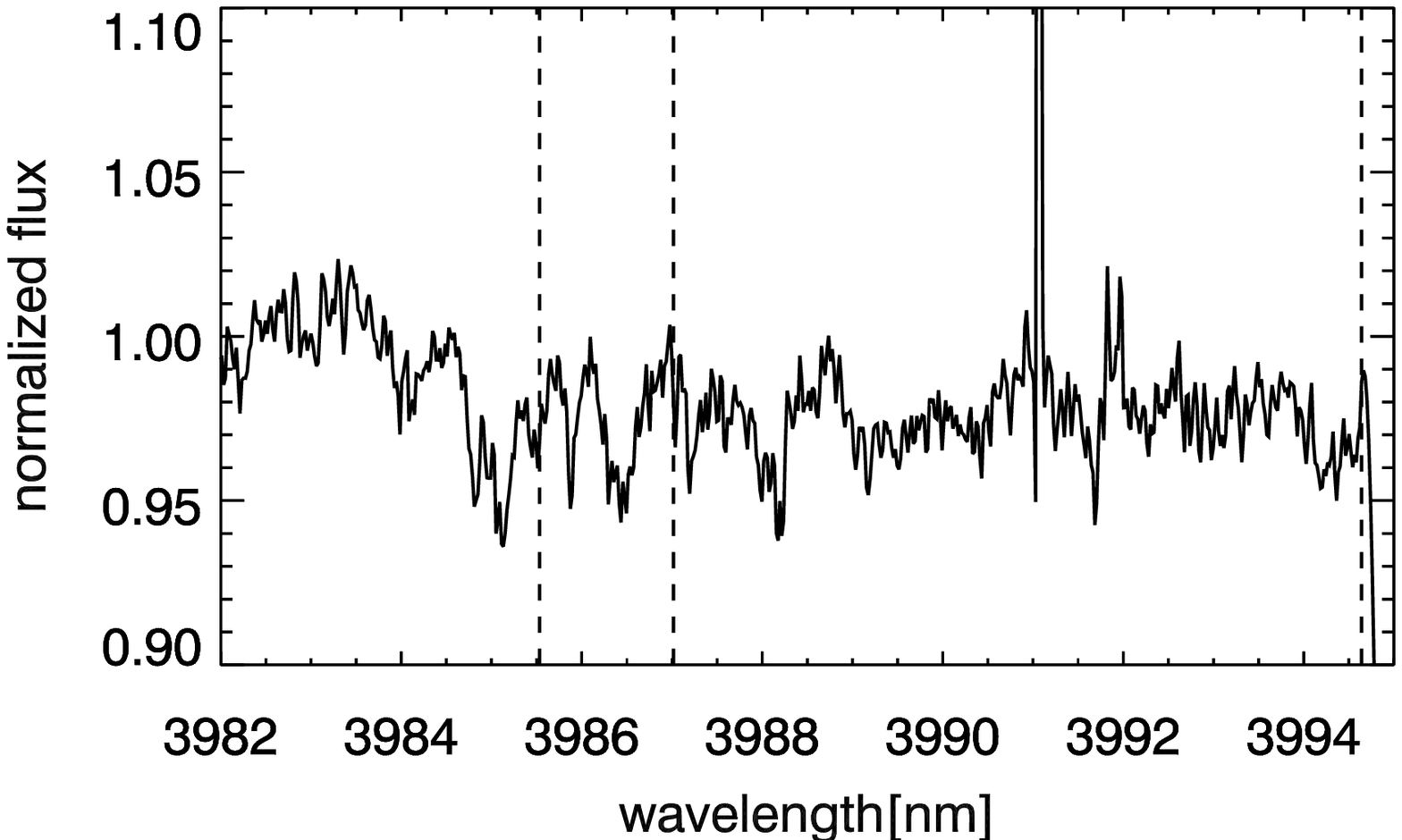}
   
      \includegraphics[width=0.42\textwidth]{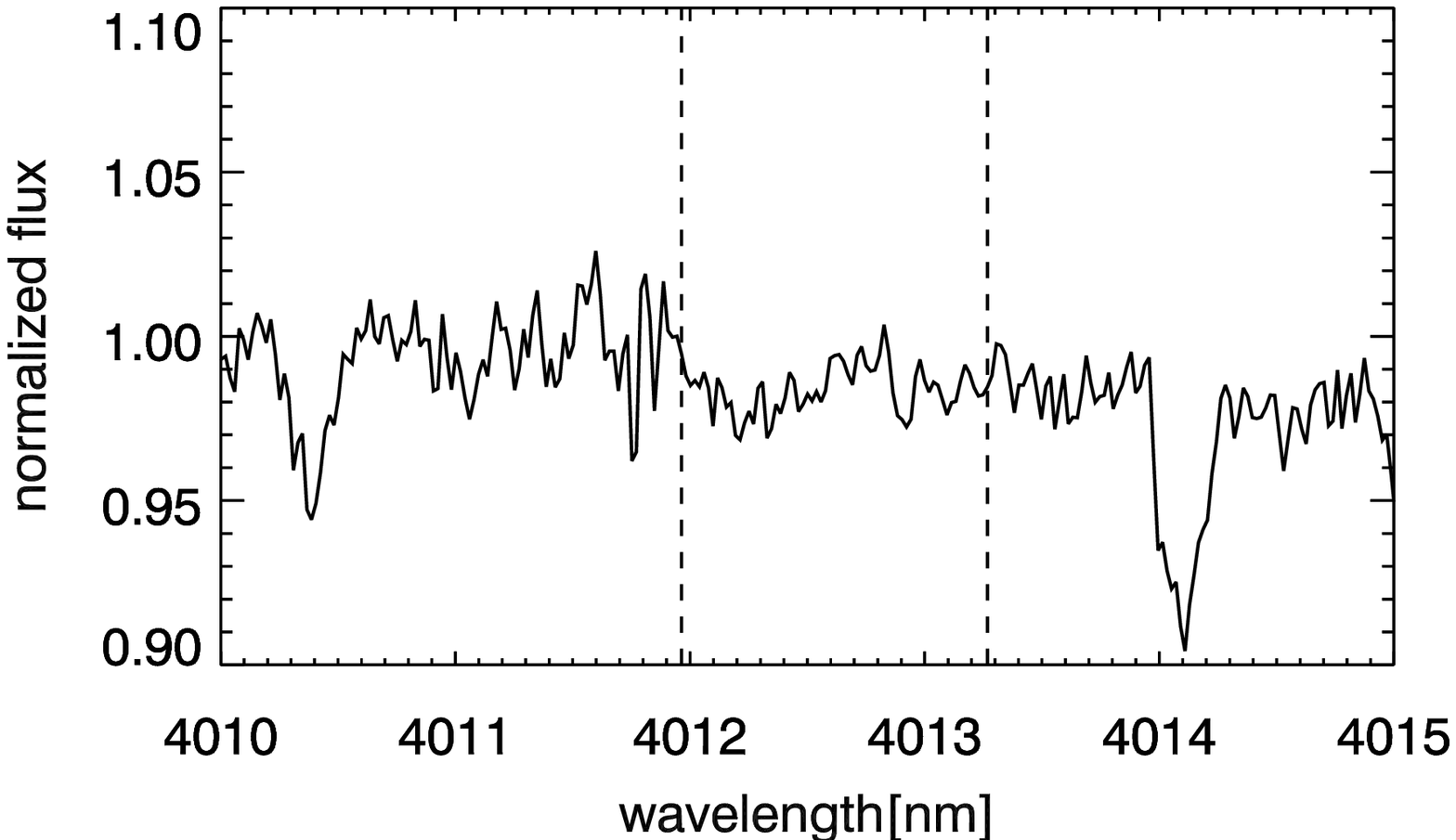}
   \includegraphics[width=0.42\textwidth]{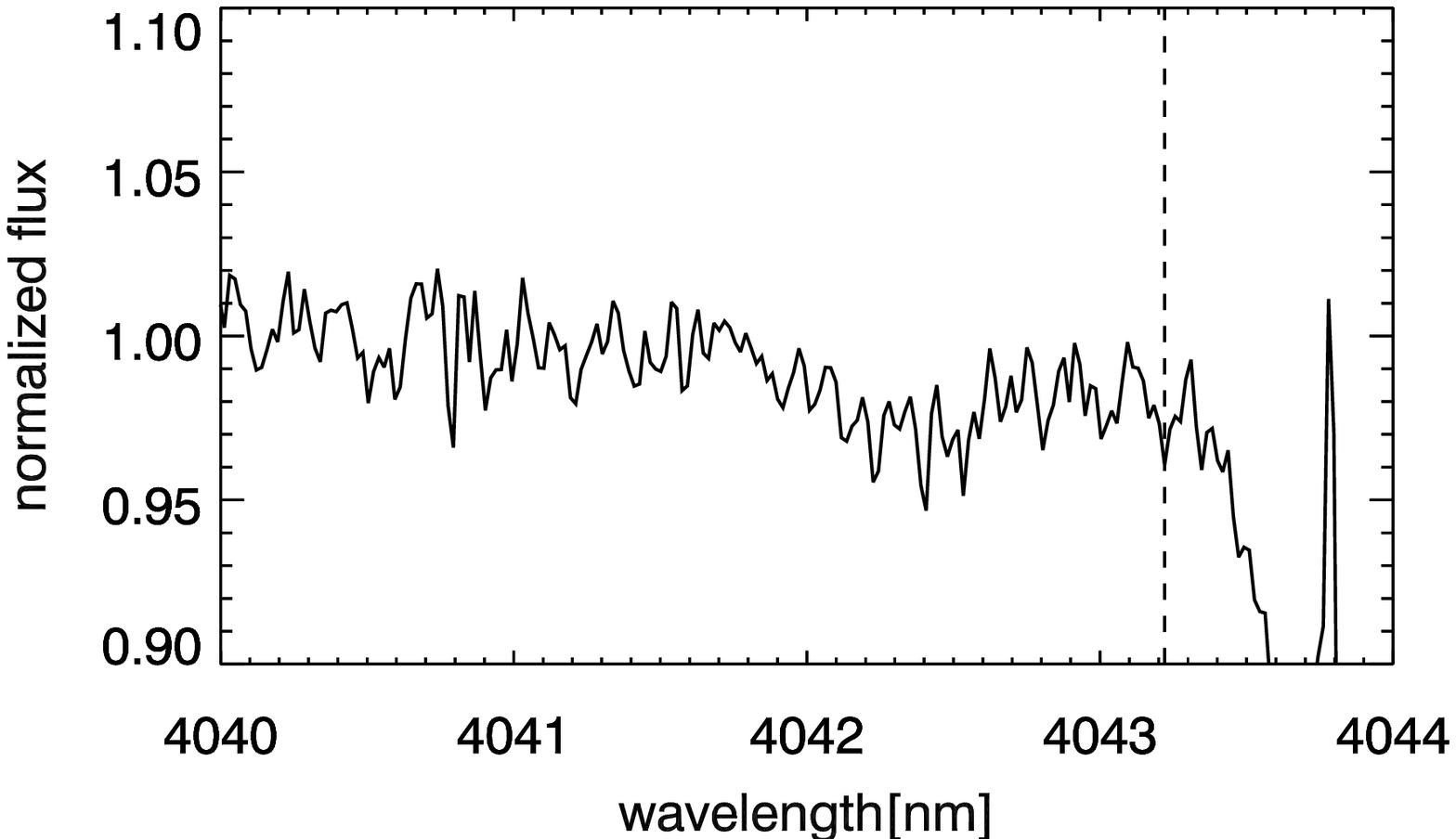}
       \caption{Spectra of HD\,209458, a zoom-in for all four detectors is shown. All observed combined light spectra, were shifted to their apparent radial velocity at the time of observation and then added up binwise. The dashed lines indicate the positions
where the H$_{3}^{+}$ emission lines are expected. }
       \label{saspec}
   \end{figure*}

The resulting  shift and add frame is shown in Fig.~\ref{saspec}. The H$_{3}^{+}$ emission line positions from \cite{Maillard1990} are highlighted by the dashed lines.  Insufficient correction of telluric lines by the modeling and the application of the shift and add technique cause artificial, broad absorption lines in the resulting frames. Since we do not apply any correction  for the crack on detector No.~2 for this part of the analysis, the crack is still visible as a large peak around 3991~nm in the resulting spectrum.
In the results from the  shift and add procedure,  we cannot detect any signs of emission peaks at the expected H$_{3}^{+}$ emission line positions. 

As a last test for the shift and add results, we create an artificial spectrum; this articifial spectrum contains only Gaussians at the positions of the H$_{3}^{+}$ line positions from \cite{Maillard1990} and a line width calculated from the thermal width of $0.052~\mathrm{nm}$ folded with the instrumental profile and line strengths of $3\sigma$, which we measure in the resulting shift and add frame at the position of the emission line at $3953.0\,\mathrm{nm}$. 
We perform a cross-correlation of this artificial spectrum with the resulting shift and add frame as a final test for hidden planetary emission signals for each detector separately. The cross-correlations do not show a significant peak.

We also search the spectra for a drop of the total flux during secondary eclipse, but  cannot identify any variances in the integrated flux that would indicate the end of the secondary eclipse.  Variations of the flux due to atmospheric distortions are too large to allow for the measurement of the $\propto 0.1\%$ secondary eclipse depth \cite[]{Diamond-Lowe2014}.

\section{Results}
\label{sec:res}
Our search for planetary emission signatures of H$_{3}^{+}$ in our observed spectra of HD~209458 was not successful. The direct search for the strongest emission lines detected in the aurora of Jupiter by \cite{Maillard1990} shows no emission peaks at the expected line positions. 
A drawback of this method is that the spectra have to be treated separately, which limits the possible signal-to-noise ratio that could be achieved by combining the observed spectra. 
Cross-correlating the spectra, after subtracting the last observed eclipse spectrum, gives signals that scatter around $0\,\mathrm{km s^{-1}}$. This method is independent of the knowledge of the precise position of the H$_{3}^{+}$ lines. A rough estimation of the differential planetary radial velocity would be sufficient to search for the radial velocity shift that would occur in case of a detection. There are weaker H$_{3}^{+}$ lines measured in the aurora of Jupiter around the strong lines we chose to search for with our direct search \cite[]{Maillard1990}. These weaker lines could help to detect a planetary signal with the cross-correlation method. 
The shift and add results show no emission features at the expected line positions. The method could retrieve a signal that was hidden in the stellar signal, since shifting the spectra with respect to the planetary radial velocity blurs the stellar lines. However, it requires the knowledge of the planetary radial velocity and ephemeris to shift the spectra precisely to the planetary radial velocity.

\subsection{ Emission limits from the  shift and add results}
\citet{Maillard1990} find the H$_{3}^{+}$ emission line at  3985.5\,nm, the $Q(3,0)$-transition, to be the strongest in their measurements for the auroral zones of Jupiter. \citet{Shkolnik2006} use a different emission line for their search for auroral H$_{3}^{+}$ emission, the  $Q(1,0)$-transition at 3953.0\,nm. 
We use our shift  and add results to estimate an upper emission limit for planetary H$_{3}^{+}$ emission from our observations of HD\,209458  (Fig.\ref{saspec}) for both  H$_{3}^{+}$ transitions. 
For the following emission limit calculations, we assume that all planetary H$_{3}^{+}$ energy is stored in one emission line, following the approach of \cite{Shkolnik2006}. 
 
We measure the standard deviation  at the positions of the H$_{3}^{+}$ emission lines at $3985.5\,\mathrm{nm}$ and $3953.0\,\mathrm{nm}$  in the resulting shift and add spectrum (See Tab.\,\ref{noisetable}). 
To estimate the full width at half maximum (FWHM)  for the emission lines, we calculate the Doppler broadening of the lines assuming a temperature of 1000~K as a rough estimate for the temperature in the lower thermosphere. H$_{3}^{+}$ is in strong emission in this temperature range \cite[]{Miller2000}.
We account for the instrumental profile, derived from the measurement of the FWHM of the telluric lines and calculate resulting line widths of  FWHM$=0.76\AA$ for the line at  $3953.0\,\mathrm{nm}$ and FWHM$=0.69\AA$ for  $3985.5\,\mathrm{nm}$.

To estimate the power output of HD~209458 in the $L'$ window, we convert the $K$ magnitude, taken from the 2MASS
 catalog \cite[]{2003yCat.2246....0C} to the $\mathrm{L}'$ magnitude using the intrinsic colors from \cite{Bessel1988}.  We calculate a $\mathrm{L}'$ magnitude of 6.258. With the $\mathrm{L}'$ magnitude, the bandwidth ($0.65\,\mu \mathrm{m}$) and flux density for Vega of the $L'$~band of $5.267\cdot 10 ^{-11} \, \mathrm{Wm^{-2}\mu m^{-1}}$ from \cite{AllensAQ}, we derive an  $\mathrm{L}'$ band flux for HD~209458 of $1.07\cdot 10 ^{-13} \,\mathrm{Wm^{-2}}$. Using the distance of $d=47.1$~pc (see Tab.~\ref{tab:systemparameters}), we derive the total $L'$ luminosity of the star to be $2.85 \cdot 10^{24}\,\mathrm{W}$. 
 This is in good agreement with the bolometric luminosity of HD~209458 by \cite{Cody2002} of $6.1\cdot 10^{26}\,\mathrm{W}$. 
For the calculation of the detection limit of our measurements, we assume that a Gaussian of $3\sigma$ peak height could be detected and derive the detection limit from the fraction of the total $\mathrm{L}'$ luminosity in the estimated H$_{3}^{+}$ equivalent width. We calculate detection limits of $5.34\cdot 10^{18}\,\mathrm{W}$ for the emission line at $3985.5\,\mathrm{nm}$  and $8.32\cdot10^{18}\,\mathrm{W}$ at $3953.0\,\mathrm{nm}$.

\begin{table}
\caption{Detection limit results.}             
\label{noisetable}      
\centering                          
\begin{tabular}{l c c c c}        
\hline\hline                 
H$_{3}^{+}$ line         &  Transition & FWHM   &  $\sigma$ &  Det. limit   \\
 $\mathrm{[nm]}$        &                       &  $\mathrm{[nm]}$ &               &  $\mathrm{[W]}$ \\  \hline
  $3953.0$              &    $Q(1,0)$    &  $0.076$ &   $0.0068$ &  $8.32\cdot 10^{18}$\\    
  $ 3985.5$      &  $Q(3,0)$     & $0.069$ &  $0.0072$   & $5.34\cdot10^{18}$ \\
\hline                                   
\end{tabular}
\end{table}

We compare our results to the theoretical H$_{3}^{+}$ limits derived by \cite{Miller2000},  \cite{Yelle2004},  and \cite{Koskinen2007}. 
\cite{Miller2000} used the Jovian ionosphere model (JIM) by \cite{Achilleos1998} and placed Jupiter in a close orbit around the sun, calculating a total H$_{3}^{+}$ output of $10^{17}\,\mathrm{W}$. \cite{Yelle2004} derived an emission limit of  $10^{16}\,\mathrm{W}$, based on a one-dimensional model of an EGP while choosing the system parameter to match those of HD~209458 and its orbiting planet.
\cite{Koskinen2007} used three-dimensional, self-consistent global simulations  of a coupled thermosphere-ionosphere model resulting in a  total power output of up to $\sim 10^{15}\,\mathrm{W}$.
\cite{Koskinen2007} expected the spectral line output for the $Q(3,0)$-transition to be $1\%$ of the total output power of H$_{3}^{+}$, hence a limit of $\sim 10^{12}\,\mathrm{W}$. The calculated emission limit for H$_{3}^{+}$ from the atmosphere of HD~209458~b from our observations is 1 to 3 orders of magnitude less sensitive than the theoretical predictions. The calculated spectral line output for the emission line at $3953.0\,\mathrm{nm}$ by \cite{Koskinen2007} is even 6 orders of magnitude away from our measurements.

Next, we compare our results to the  H$_{3}^{+}$ emission limits reported by  \cite{Shkolnik2006} and \cite{Laughlin2008}.  
\cite{Shkolnik2006} measured detection limits for six hot Jupiter systems of different spectral types and reported a detection limit of $6.3\cdot10^{17}\,\mathrm{W}$ at 3953.0\,nm for the M-dwarf GJ~436 that is one magnitude lower than our measurements. This difference is mainly because of its late spectral type, as the data quality is similar for all spectra used for their analysis. 
The detection limit results for the late F dwarfs in their sample are comparable to our limit measured for HD~209458~b, a G0 star.  
 \cite{Laughlin2008} reported an H$_{3}^{+}$  limit for  $\tau$-Boo of $9.0\cdot10^{17}\,\mathrm{W,}$ which is also half a magnitude better than our measurement.
Considering the much higher resolution of CRIRES in comparison to CSHELL, but the small amount of observing time for our data, achieving roughly the same order of magnitude for the emission limits seems realistic.
 Our data was obtained in very short observation time and bad weather. However, even optimal conditions would require significant effort to push the detection limits into the regime of the model predictions as we show in the next Section.

\subsection{Prospects for future observations}
Our upper limits for HD~209458  were achieved in a total integration time of only 600 seconds and with bad weather conditions.  
There are hot Jupiter hosting stars that are closer to Earth, which increases the detectability of planetary emission. We estimate that for a system at 10pc distance one could obtain data with sufficient quality to push the detection limit to $5 \cdot 10^{16}\,\mathrm{W}$ in about six hours observation time with CRIRES at the VLT. The data quality would allow us to reach below the theoretical limit given by \cite{Miller2000} and into the regime of the limit given by \cite{Yelle2004} for a few close star-planet systems. Even though CRIRES is the most promising instrument available for this task, it would still not be sufficient to descend to  the estimations of \cite{Koskinen2007}. 
The upgrade of CRIRES to CRIRES+ \cite[]{criresplus}, enabling the spectrograph for cross-dispersion, will promote the search for H$_{3}^{+}$. The continuous spectrum of CRIRES+ will enable the observation of many H$_{3}^{+}$ emission line positions and thus cross-correlation techniques will be more promising compared to the four separate detectors of CRIRES. 

Future telescopes with larger diameters will allow for  observations that reach the theoretical predictions of H$_{3}^{+}$ of exoplanet atmospheres. The European Extremely Large Telescope\footnote{\url{http://www.eso.org/sci/facilities/eelt/docs/e-elt_constrproposal.pdf}} (E-ELT) will have a diameter of 39~m that will dramatically reduce the required exposure times. In Table~\ref{tab:VLTEELT} we compare the exposure times for the observation of H$_{3}^{+}$ in an exemplary hot Jupiter system. For these estimations we calculate a star-planet-contrast from the theoretical planetary emission limits and use a blackbody for the stellar emission at the H$_{3}^{+}$ line position at $3953.0\,\mathrm{nm}$. We use the star-planet-contrast to obtain an estimate for the  required signal-to-noise ratio for a detection of a planetary signal at this line position.
As an exemplary planetary system for the calculation of the observation times, we choose the Gl~86 system, which hosts a $4.01~\mathrm{M}_{\mathrm{Jup}}$ planet at $0.11~\mathrm{AU}$ orbital distance \cite[]{Queloz2000-Gl86}. The stellar system is only $10.9~\mathrm{pc}$ away, which increases the possibility of detecting H$_{3}^{+}$ emission from Earth. Candidate systems for the detection of H$_{3}^{+}$ emission should preferably be tidally locked, so that the expected emission lines are narrow and very high resolution spectroscopy would enhance the ratio of planetary emission to stellar continuum flux. 
 \begin{table}
\caption{Comparison of exposure times for different H$_{3}^{+}$ emission limits  with  VLT and E-ELT. We chose Gl~86 as an example candidate system.}             
\label{tab:VLTEELT}      
\centering                          
\begin{tabular}{l  r r}        
\hline\hline                 
  Emission Limit        & Exp. Time VLT         & Exp. Time E-ELT  \\   
         $\mathrm{[W]}$ &                [min]  & [min]  \\   
         $10^{17}$      &        94                     &  4  \\   
 $5\cdot10^{16}$        &       370             &  16\\
        $10^{16}$       &  9300                 & 390 \\
        $10^{15}$       &970000         & 40000 \\
\hline                                   
\end{tabular}
\end{table} 

 We calculate the exposure times for the VLT observations with the CRIRES exposure time calculator\footnote{\url{http://www.eso.org/observing/etc/bin/gen/form?INS.NAME=CRIRES+INS.MODE=swspectr}} and derive the expected exposure times for the E-ELT assuming a similar performing instrument as CRIRES and accounting for the larger collecting area. 
 Collecting spectra of hot Jupiter systems with a quality sufficient to test the predictions by \cite{Miller2000} and \cite{Yelle2004} will very well be possible. 
However,  reaching the emission limits predicted by \cite{Koskinen2007} for a single emission line of $10^{12}\,\mathrm{W}$ with a ground-based instrument seems unlikely.
Evidently, H$_{3}^{+}$ rotational-vibrational transitions would best be measured from space. MIRI\footnote{\url{http://www.jwst.nasa.gov/miri.html}} at James Webb Space Telescope however, does not provide for spectroscopy in this regime. Hence, an instrument such as METIS\footnote{\url{http://metis.strw.leidenuniv.nl/}} on the E-ELT will be of great value \cite[]{2014SPIE.9147E..21B}. As presented right now, METIS Echelle spectroscopy is implemented as a single order integral field mode. For this project a classical cross-dispersed spectrograph such as the updated CRIRES would be preferred.


\section{Summary}
\label{sec:sum}
We carried out an observational test for the detection of H$_{3}^{+}$ in the atmosphere of a hot Jupiter by analyzing spectra of HD\,209458 obtained with CRIRES at the VLT. Our dataset contained 12 spectra taken during secondary eclipse of the transiting planet HD\,209458\,b and four spectra that contained the combined light of the star and (possibly) the  atmosphere of the planet.

We searched for H$_{3}^{+}$  emission at those line positions that have been found in strong emission in the aurora of Jupiter by \cite{Maillard1990}. Our analysis consisted of three different approaches. First, we searched directly for the emission in the observed spectra.
Second, we tested a cross-correlation approach, and as a last test, we created a shift and add spectrum by shifting all spectra with respect to their apparent planetary radial velocity at the time of observation and adding up the flux binwise. We could not identify any signs of  H$_{3}^{+}$ emission in any of the three steps of the analysis.

From the shift and add results, we calculated an upper emission limit for the two H$_{3}^{+}$ emission lines that are reported to have the most flux in this wavelength region in the  auroral zones of Jupiter. We calculate the emission limit for the line at  3953.0\,nm [$Q(1,0)$]  to be $8.32\cdot10^{18}\,\mathrm{W}$   and a limit of f $5.34\cdot 10^{18}\,\mathrm{W}$ for the emission line at 3985.5\,nm [$Q(3,0)$].
Our emission limits are 1 to 3 orders of magnitudes above theoretical predictions in the literature.
 
We estimated that data quality sufficient to reach the predicted emission limit of $\sim 10^{17}\,\mathrm{W}$ by \cite{Miller2000} and even a detection  limit of  $5 \cdot 10^{16}\,\mathrm{W}$ can be achieved with CRIRES for star-planet systems that are close to Earth.
However, the faintest  H$_{3}^{+}$ emission limit for a single line flux of  $\sim 10^{12}\,\mathrm{W}$ for a close-in hot Jupiter derived by  \cite{Koskinen2007} will be difficult to reach with ground-based facilities in the near future. Future giant telescopes, such as the E-ELT will dramatically reduce the required exposure times and thus enable a large sample search for  H$_{3}^{+}$ emission from hot Jupiter atmospheres and may very well push the detection limit down to $10^{15 }\,\mathrm{W}$.
      
\begin{acknowledgements}
We warmly thank Ulf Seemann and Sebastian Sch\"afer for their help with the data handling and reduction and Johanna Kerch for a thorough reading of the manuscript.
Lea Lenz acknowledges financial support from the Deutsche Forschungsgemeinschaft under DFG GrK 1351. Ansgar Reiners has received research funding as a Heisenberg Professor under DFG 1664/9-1.  \end{acknowledgements}

\bibliographystyle{aa}
\bibliography{references}

\begin{thebibliography}{33}
\expandafter\ifx\csname natexlab\endcsname\relax\def\natexlab#1{#1}\fi

\bibitem[{{Achilleos} {et~al.}(1998){Achilleos}, {Miller}, {Tennyson},
  {Aylward}, {Mueller-Wodarg}, \& {Rees}}]{Achilleos1998}
{Achilleos}, N., {Miller}, S., {Tennyson}, J., {et~al.} 1998, \jgr, 103, 20089

\bibitem[{{Beaulieu} {et~al.}(2010){Beaulieu}, {Kipping}, {Batista}, {Tinetti},
  {Ribas}, {Carey}, {Noriega-Crespo}, {Griffith}, {Campanella}, {Dong},
  {Tennyson}, {Barber}, {Deroo}, {Fossey}, {Liang}, {Swain}, {Yung}, \&
  {Allard}}]{Beaulieu2010}
{Beaulieu}, J.~P., {Kipping}, D.~M., {Batista}, V., {et~al.} 2010, \mnras, 409,
  963

\bibitem[{{Bessell} \& {Brett}(1988)}]{Bessel1988}
{Bessell}, M.~S. \& {Brett}, J.~M. 1988, \pasp, 100, 1134

\bibitem[{{Brandl} {et~al.}(2014){Brandl}, {Feldt}, {Glasse}, {Guedel},
  {Heikamp}, {Kenworthy}, {Lenzen}, {Meyer}, {Molster}, {Paalvast}, {Pantin},
  {Quanz}, {Schmalzl}, {Stuik}, {Venema}, \& {Waelkens}}]{2014SPIE.9147E..21B}
{Brandl}, B.~R., {Feldt}, M., {Glasse}, A., {et~al.} 2014, in Society of
  Photo-Optical Instrumentation Engineers (SPIE) Conference Series, Vol. 9147,
  Society of Photo-Optical Instrumentation Engineers (SPIE) Conference Series,
  21

\bibitem[{{Brittain} \& {Rettig}(2002)}]{Brittain2002}
{Brittain}, S.~D. \& {Rettig}, T.~W. 2002, \nat, 418, 57

\bibitem[{{Charbonneau} {et~al.}(2002){Charbonneau}, {Brown}, {Noyes}, \&
  {Gilliland}}]{Charbonneau2002}
{Charbonneau}, D., {Brown}, T.~M., {Noyes}, R.~W., \& {Gilliland}, R.~L. 2002,
  \apj, 568, 377

\bibitem[{{Cody} \& {Sasselov}(2002)}]{Cody2002}
{Cody}, A.~M. \& {Sasselov}, D.~D. 2002, \apj, 569, 451

\bibitem[{{Cox}(2000)}]{AllensAQ}
{Cox}, A.~N. 2000, {Allen's astrophysical quantities}

\bibitem[{{Cutri} {et~al.}(2003){Cutri}, {Skrutskie}, {van Dyk}, {Beichman},
  {Carpenter}, {Chester}, {Cambresy}, {Evans}, {Fowler}, {Gizis}, {Howard},
  {Huchra}, {Jarrett}, {Kopan}, {Kirkpatrick}, {Light}, {Marsh}, {McCallon},
  {Schneider}, {Stiening}, {Sykes}, {Weinberg}, {Wheaton}, {Wheelock}, \&
  {Zacarias}}]{2003yCat.2246....0C}
{Cutri}, R.~M., {Skrutskie}, M.~F., {van Dyk}, S., {et~al.} 2003, VizieR Online
  Data Catalog, 2246, 0

\bibitem[{{Diamond-Lowe} {et~al.}(2014){Diamond-Lowe}, {Stevenson}, {Bean},
  {Line}, \& {Fortney}}]{Diamond-Lowe2014}
{Diamond-Lowe}, H., {Stevenson}, K.~B., {Bean}, J.~L., {Line}, M.~R., \&
  {Fortney}, J.~J. 2014, \apj, 796, 66

\bibitem[{{Drossart} {et~al.}(1989){Drossart}, {Maillard}, {Caldwell}, {Kim},
  {Watson}, {Majewski}, {Tennyson}, {Miller}, {Atreya}, {Clarke}, {Waite}, \&
  {Wagener}}]{Drossart1989}
{Drossart}, P., {Maillard}, J.-P., {Caldwell}, J., {et~al.} 1989, \nat, 340,
  539

\bibitem[{{Geballe} {et~al.}(1993){Geballe}, {Jagod}, \& {Oka}}]{Geballe1993}
{Geballe}, T.~R., {Jagod}, M.-F., \& {Oka}, T. 1993, \apjl, 408, L109

\bibitem[{{Goto} {et~al.}(2005){Goto}, {Geballe}, {McCall}, {Usuda}, {Suto},
  {Terada}, {Kobayashi}, \& {Oka}}]{Goto2005}
{Goto}, M., {Geballe}, T.~R., {McCall}, B.~J., {et~al.} 2005, \apj, 629, 865

\bibitem[{{Kaeufl} {et~al.}(2004){Kaeufl}, {Ballester}, {Biereichel},
  {Delabre}, {Donaldson}, {Dorn}, {Fedrigo}, {Finger}, {Fischer}, {Franza},
  {Gojak}, {Huster}, {Jung}, {Lizon}, {Mehrgan}, {Meyer}, {Moorwood}, {Pirard},
  {Paufique}, {Pozna}, {Siebenmorgen}, {Silber}, {Stegmeier}, \&
  {Wegerer}}]{2004SPIE.5492.1218K}
{Kaeufl}, H.-U., {Ballester}, P., {Biereichel}, P., {et~al.} 2004, in Society
  of Photo-Optical Instrumentation Engineers (SPIE) Conference Series, Vol.
  5492, Society of Photo-Optical Instrumentation Engineers (SPIE) Conference
  Series, ed. A.~F.~M. {Moorwood} \& M.~{Iye}, 1218--1227

\bibitem[{{Knutson} {et~al.}(2007){Knutson}, {Charbonneau}, {Noyes}, {Brown},
  \& {Gilliland}}]{Knutson2007}
{Knutson}, H.~A., {Charbonneau}, D., {Noyes}, R.~W., {Brown}, T.~M., \&
  {Gilliland}, R.~L. 2007, \apj, 655, 564

\bibitem[{{Koskinen} {et~al.}(2007){Koskinen}, {Aylward}, {Smith}, \&
  {Miller}}]{Koskinen2007}
{Koskinen}, T.~T., {Aylward}, A.~D., {Smith}, C.~G.~A., \& {Miller}, S. 2007,
  \apj, 661, 515

\bibitem[{{Koskinen} {et~al.}(2010){Koskinen}, {Yelle}, {Lavvas}, \&
  {Lewis}}]{Koskinen2010a}
{Koskinen}, T.~T., {Yelle}, R.~V., {Lavvas}, P., \& {Lewis}, N.~K. 2010, \apj,
  723, 116

\bibitem[{{Lam} {et~al.}(1997){Lam}, {Achilleos}, {Miller}, {Tennyson},
  {Trafton}, {Geballe}, \& {Ballester}}]{Lam1997}
{Lam}, H.~A., {Achilleos}, N., {Miller}, S., {et~al.} 1997, \icarus, 127, 379

\bibitem[{{Lammer} {et~al.}(2003){Lammer}, {Selsis}, {Ribas}, {Guinan},
  {Bauer}, \& {Weiss}}]{Lammer2003}
{Lammer}, H., {Selsis}, F., {Ribas}, I., {et~al.} 2003, \apjl, 598, L121

\bibitem[{{Laughlin} {et~al.}(2008){Laughlin}, {Troutman}, {Brittain}, \&
  {Rettig}}]{Laughlin2008}
{Laughlin}, L., {Troutman}, M.~R., {Brittain}, S., \& {Rettig}, T.~W. 2008, in
  Bulletin of the American Astronomical Society, Vol.~40, American Astronomical
  Society Meeting Abstracts \#212, 529

\bibitem[{{Lebzelter} {et~al.}(2012){Lebzelter}, {Seifahrt}, {Uttenthaler},
  {Ramsay}, {Hartman}, {Nieva}, {Przybilla}, {Smette}, {Wahlgren}, {Wolff},
  {Hussain}, {K{\"a}ufl}, \& {Seemann}}]{Lebzelter2012}
{Lebzelter}, T., {Seifahrt}, A., {Uttenthaler}, S., {et~al.} 2012, \aap, 539,
  A109

\bibitem[{{Maillard} \& {Miller}(2011)}]{Maillard2011}
{Maillard}, J. \& {Miller}, S. 2011, in Astronomical Society of the Pacific
  Conference Series, Vol. 450, Molecules in the Atmospheres of Extrasolar
  Planets, ed. J.~P. {Beaulieu}, S.~{Dieters}, \& G.~{Tinetti}, 19--21

\bibitem[{{Maillard} {et~al.}(1990){Maillard}, {Drossart}, {Watson}, {Kim}, \&
  {Caldwell}}]{Maillard1990}
{Maillard}, J.-P., {Drossart}, P., {Watson}, J.~K.~G., {Kim}, S.~J., \&
  {Caldwell}, J. 1990, \apjl, 363, L37

\bibitem[{{Miller} {et~al.}(2000){Miller}, {Achilleos}, {Ballester}, {Geballe},
  {Joseph}, {Prang{\'e}}, {Rego}, {Stallard}, {Tennyson}, {Trafton}, \&
  {Waite}}]{Miller2000}
{Miller}, S., {Achilleos}, N., {Ballester}, G.~E., {et~al.} 2000, in Royal
  Society of London Philosophical Transactions Series A, Vol. 358, Astronomy,
  physics and chemistry of H$^{+}$$_{3}$, 2485

\bibitem[{{Neale} {et~al.}(1996){Neale}, {Miller}, \& {Tennyson}}]{Neale1996}
{Neale}, L., {Miller}, S., \& {Tennyson}, J. 1996, \apj, 464, 516

\bibitem[{{Oliva} {et~al.}(2014){Oliva}, {Tozzi}, {Ferruzzi}, {Origlia},
  {Hatzes}, {Follert}, {Loewinger}, {Piskunov}, {Heiter}, {Lockhart},
  {Marquart}, {Stempels}, {Reiners}, {Anglada-Escude}, {Seemann}, {Dorn},
  {Bristow}, {Baade}, {Delabre}, {Gojak}, {Grunhut}, {Klein}, {Hilker}, {Ives},
  {Jung}, {Kaeufl}, {Kerber}, {Lizon}, {Pasquini}, {Paufique}, {Pozna},
  {Smette}, {Smoker}, \& {Valenti}}]{criresplus}
{Oliva}, E., {Tozzi}, A., {Ferruzzi}, D., {et~al.} 2014, ArXiv e-prints

\bibitem[{{Queloz} {et~al.}(2000){Queloz}, {Mayor}, {Weber}, {Bl{\'e}cha},
  {Burnet}, {Confino}, {Naef}, {Pepe}, {Santos}, \& {Udry}}]{Queloz2000-Gl86}
{Queloz}, D., {Mayor}, M., {Weber}, L., {et~al.} 2000, \aap, 354, 99

\bibitem[{{Seifahrt} {et~al.}(2010){Seifahrt}, {K{\"a}ufl}, {Z{\"a}ngl},
  {Bean}, {Richter}, \& {Siebenmorgen}}]{Seifahrt2010}
{Seifahrt}, A., {K{\"a}ufl}, H.~U., {Z{\"a}ngl}, G., {et~al.} 2010, \aap, 524,
  A11

\bibitem[{{Shkolnik} {et~al.}(2006){Shkolnik}, {Gaidos}, \&
  {Moskovitz}}]{Shkolnik2006}
{Shkolnik}, E., {Gaidos}, E., \& {Moskovitz}, N. 2006, \aj, 132, 1267

\bibitem[{{Southworth}(2010)}]{Southworth2010}
{Southworth}, J. 2010, \mnras, 408, 1689

\bibitem[{{Trafton} {et~al.}(1993){Trafton}, {Geballe}, {Miller}, {Tennyson},
  \& {Ballester}}]{Trafton1993}
{Trafton}, L.~M., {Geballe}, T.~R., {Miller}, S., {Tennyson}, J., \&
  {Ballester}, G.~E. 1993, \apj, 405, 761

\bibitem[{{Wittenmyer} {et~al.}(2005){Wittenmyer}, {Welsh}, {Orosz}, {Schultz},
  {Kinzel}, {Kochte}, {Bruhweiler}, {Bennum}, {Henry}, {Marcy}, {Fischer},
  {Butler}, \& {Vogt}}]{Wittenmyer2005}
{Wittenmyer}, R.~A., {Welsh}, W.~F., {Orosz}, J.~A., {et~al.} 2005, \apj, 632,
  1157

\bibitem[{{Yelle}(2004)}]{Yelle2004}
{Yelle}, R.~V. 2004, \icarus, 170, 167

\end{thebibliography}


\end{document}